\def\year{2018}\relax
\documentclass[letterpaper]{article} %DO NOT CHANGE THIS
\usepackage{aaai18}  %Required
\usepackage{times}  %Required
\usepackage{helvet}  %Required
\usepackage{courier}  %Required
\usepackage{url}  %Required
\usepackage{graphicx}  %Required
\usepackage{latexsym}
\usepackage{booktabs} % For formal tables
\usepackage{graphicx}
\usepackage{algorithm}
\usepackage[noend]{algpseudocode} 
\usepackage{caption}
\usepackage{amsmath}
\usepackage{graphics}
\usepackage{float}
\usepackage{xcolor}
\usepackage{etoolbox}
\usepackage{tikz}
\usepackage{paralist}
\usepackage{subcaption}

\begin{document}
\title{Analyzing Gender Stereotyping in Bollywood Movies}
% \author{AAAI PAnalyzing Gender Stereotyping in Bollywood Moviesress\\
% Association for the Advancement of Artificial Intelligence\\
% 2275 East Bayshore Road, Suite 160\\
% Palo Alto, California 94303\\
% }

\author{Nishtha Madaan\textsuperscript{1}, Sameep Mehta\textsuperscript{1}, Taneea S Agrawaal\textsuperscript{2}, Vrinda Malhotra\textsuperscript{2}, Aditi Aggarwal\textsuperscript{2}, Mayank Saxena\textsuperscript{3}\\ 
\textsuperscript{1}IBM Research-INDIA\\
\textsuperscript{2}IIIT-Delhi\\
\textsuperscript{3}DTU-Delhi\\
\{nishthamadaan, sameepmehta\}@in.ibm.com , \{taneea14166, vrinda14122, aditi16004\}@iiitd.ac.in, \{mayank26saxena\}@gmail.com
}

\maketitle
\begin{abstract}
The presence of gender stereotypes in many aspects of society is a well-known phenomenon. In this paper, we focus on studying such stereotypes and bias in Hindi movie industry ({\it Bollywood}). We analyze movie plots and posters for all movies released since 1970. The gender bias is detected by semantic modeling of plots at inter-sentence and intra-sentence level. Different features like occupation, introduction of cast in text, associated actions and descriptions are captured to show the pervasiveness of gender bias and stereotype in movies. We derive a semantic graph and compute centrality of each character and observe similar bias there. We also show that such bias is not applicable for movie posters where females get equal importance even though their character has little or no impact on the movie plot. Furthermore, we explore the movie trailers to estimate on-screen time for males and females and also study the portrayal of emotions by gender in them. 
% \mayank{Lastly, we analyze all movie trailers released from 2008 till 2017 and show that men get twice as much as on-screen time as women. We also see that the portrayal of males and females differs quite a lot with respect to the emotions exhibited by them.} 
The silver lining is that our system was able to identify 30 movies over last 3 years where such stereotypes were broken.
\end{abstract}
 
\noindent 
\section{Introduction}
Movies are a reflection of the society. They mirror (with creative liberties) the problems, issues, thinking \& perception of the contemporary society. Therefore, we believe movies could act as the proxy to understand how prevalent gender bias and stereotypes are in any society. In this paper, we leverage NLP and image understanding techniques to quantitatively study this bias. To further motivate the problem we pick a small section from the plot of a blockbuster movie.

\emph{"Rohit is an aspiring singer who works as a salesman in a car showroom, run by Malik (Dalip Tahil). One day he meets Sonia Saxena (Ameesha Patel), daughter of Mr. Saxena (Anupam Kher), when he goes to deliver a car to her home as her birthday present."}

This piece of text is taken from the \emph{plot} of Bollywood movie \emph{Kaho Na Pyaar Hai}. This simple two line plot showcases the issue in following fashion:

\begin{asparaenum}
\item Male (Rohit) is portrayed with a profession \& an aspiration
\item Male (Malik) is a business owner
\end{asparaenum}

In contrast, the female role is introduced with no profession or aspiration. The introduction, itself, is dependent upon another male character {\it "daughter of"}!

One goal of our work is to analyze and quantify gender-based stereotypes by studying the demarcation of roles designated to males and females. We measure this by performing an intra-sentence and inter-sentence level analysis of movie plots combined with the cast information. Capturing information from sentences helps us perform a holistic study of the corpus. Also, it helps us in capturing the characteristics exhibited by male and female class. We have extracted movies pages of all the Hindi movies released from 1970-present from Wikipedia. We also employ deep image analytics to capture such bias in movie posters and previews.

\subsection{Analysis Tasks}
We focus on following tasks to study gender bias in Bollywood. 
\begin{asparaenum}[I)]
    \item \textbf{Occupations and Gender Stereotypes}- How are males portrayed in their jobs vs females? How are these levels different? How does it correlate to gender bias and stereotype?
    \item \textbf{Appearance and Description} - How are males and females described on the basis of their appearance? How do the descriptions differ in both of them? How does that indicate gender stereotyping?
    \item \textbf{Centrality of Male and Female Characters} - What is the role of males and females in movie plots?  How does the amount of male being central or female being central differ? How does it present a male or female bias?
    \item \textbf{Mentions(Image vs Plot)} - How many males and females are the faces of the promotional posters? How does this correlate to them being mentioned in the plot? What results are conveyed on the combined analysis?
    \item \textbf{Dialogues} - How do the number of dialogues differ between a male cast and a female cast in official movie script?
    \item \textbf{Singers} - Does the same bias occur in movie songs? How does the distribution of singers with gender vary over a period of time for different movies?
    \item \textbf{Female-centric Movies}- Are the movie stories and portrayal of females evolving? Have we seen female-centric movies in the recent past?
    \item \textbf{Screen Time} - Which gender, if any, has a greater screen time in movie trailers?
\item \textbf{Emotions of Males and Females} - Which emotions are most commonly displayed by males and females in a movie trailer? Does this correspond with the gender stereotypes which exist in society?
\end{asparaenum}

% \mayank{In our analysis of the movie trailers we focus on the following two tasks:
% \begin{asparaenum}[I)]
% \item Screen Time - Which gender, if any, has a greater screen time in movie trailers?
% \item Portrayal of Male and Female Characters - Which emotions are most commonly displayed by males and females in a movie trailer? Does this correspond with the gender stereotypes which exist in society?
% \end{asparaenum}
% }

\section{Related Work}
While there are recent works where gender bias has been studied in different walks of life \cite{soklaridis2017gender},\cite{ macnell2015s}, \cite{carnes2015effect}, \cite{terrell2017gender}, \cite{saji2016gender}, the analysis majorly involves information retrieval tasks involving a wide variety of prior work in this area. \cite{fast2016shirtless} have worked on gender stereotypes in English fiction particularly on the Online Fiction Writing Community. The work deals primarily with the analysis of how males and females behave and are described in this online fiction. Furthermore, this work also presents that males are over-represented and finds that traditional gender stereotypes are common throughout every genre in the online fiction data used for analysis. \\ Apart from this, various works where Hollywood movies have been analyzed for having such gender bias present in them \cite{blog}. Similar analysis has been done on children books \cite{gooden2001gender} and music lyrics \cite{millar2008selective} which found that men are portrayed as strong and violent, and on the other hand, women are associated with home and are considered to be gentle and less active compared to men. These studies have been very useful to uncover the trend but the derivation of these analyses has been done on very small data sets. In some works, gender drives the decision for being hired in corporate organizations \cite{dobbin2012corporate}. Not just hiring, it has been shown that human resource professionals' decisions on whether an employee should get a raise have also been driven by gender stereotypes by putting down female claims of raise requests. While, when it comes to consideration of opinion, views of females are weighted less as compared to those of men \cite{otterbacher2015linguistic}. On social media and dating sites, women are judged by their appearance while men are judged mostly by how they behave \cite{rose2012face,otterbacher2015linguistic,fiore2008assessing}. When considering occupation, females are often designated lower level roles as compared to their male counterparts in image search results of occupations \cite{kay2015unequal}. In our work we extend these analyses for Bollywood movies.\\ 
The motivation for considering Bollywood movies is three fold: 
\begin{asparaenum}[a)]
    \item The data is very diverse in nature. Hence finding how gender stereotypes exist in this data becomes an interesting study.
    \item The data-set is large. We analyze 4000 movies which cover all the movies since 1970.  So it becomes a good first step to develop computational tools to analyze the existence of stereotypes over a period of time.
    \item These movies are a reflection of society. It is a good first step to look for such gender bias in this data so that necessary steps can be taken to remove these biases.
\end{asparaenum}

\section{Data and Experimental Study}
\subsection{Data Selection}
We deal with (three) different types of data for Bollywood Movies to perform the analysis tasks-

\subsubsection{Movies Data}
Our data-set consist of all Hindi movie pages from Wikipedia. The data-set contains 4000 movies for 1970-2017 time period. We extract movie title, cast information, plot, soundtrack information and images associated for each movie. For each listed cast member, we traverse their wiki pages to extract gender information. Cast Data consists of data for 5058 cast members who are Females and 9380 who are Males.
{\it Since we did not have access to too many official scripts, we use Wikipedia plot as proxy. We strongly believe that the Wikipedia plot represent the correct story line. If an actor had an important role in the movie, it is highly unlikely that wiki plot will miss the actor altogether.}

\subsubsection{Movies Scripts Data}
We obtained PDF scripts of 13 Bollywood movies which are available online. The PDF scripts are converted into structured HTML using \cite{ibm}. We use these HTML for our analysis tasks.

\subsubsection{Movie Preview Data}
Our data-set consists of 880 official movie trailers of movies released between 2008 and 2017. These trailers were obtained from YouTube. The mean and standard deviation of the duration of the all videos is 146 and 35 seconds respectively. The videos have a frame rate of 25 FPS and a resolution of 480p. Each 25\textsuperscript{th} frame of the video is extracted and analyzed using face classification for gender and emotion detection \cite{face_classification}.

\subsection{Task and Approach}
In this section, we discuss the tasks we perform on the movie data extracted from Wikipedia and the scripts. Further, we define the approach we adopt to perform individual tasks and then study the inferences. At a broad level, we divide our analysis in four groups. These can be categorized as follows-
\begin{asparaenum}[a)]
    \item \emph{At intra-sentence level} - We perform this analysis at a sentence level where each sentence is analyzed independently. We do not consider context in this analysis.
    \item \emph{At inter-sentence level} - We perform this analysis at a multi-sentence level where we carry context from a sentence to other and then analyze the complete information.
    \item \emph{Image and Plot Mentions} - We perform this analysis by correlating presence of genders in movie posters and in plot mentions.
    \item \emph{At Video level} - We perform this analysis by doing gender and emotion detection on the frames for each video. \cite{face_classification}
\end{asparaenum}

We define different tasks corresponding to each level of analysis.

\subsubsection{Tasks at Intra-Sentence level}
% Column wise Adjectives
\begin{figure}[t!]
\begin{subfigure}{0.25\textwidth}
 \includegraphics[width=\linewidth]{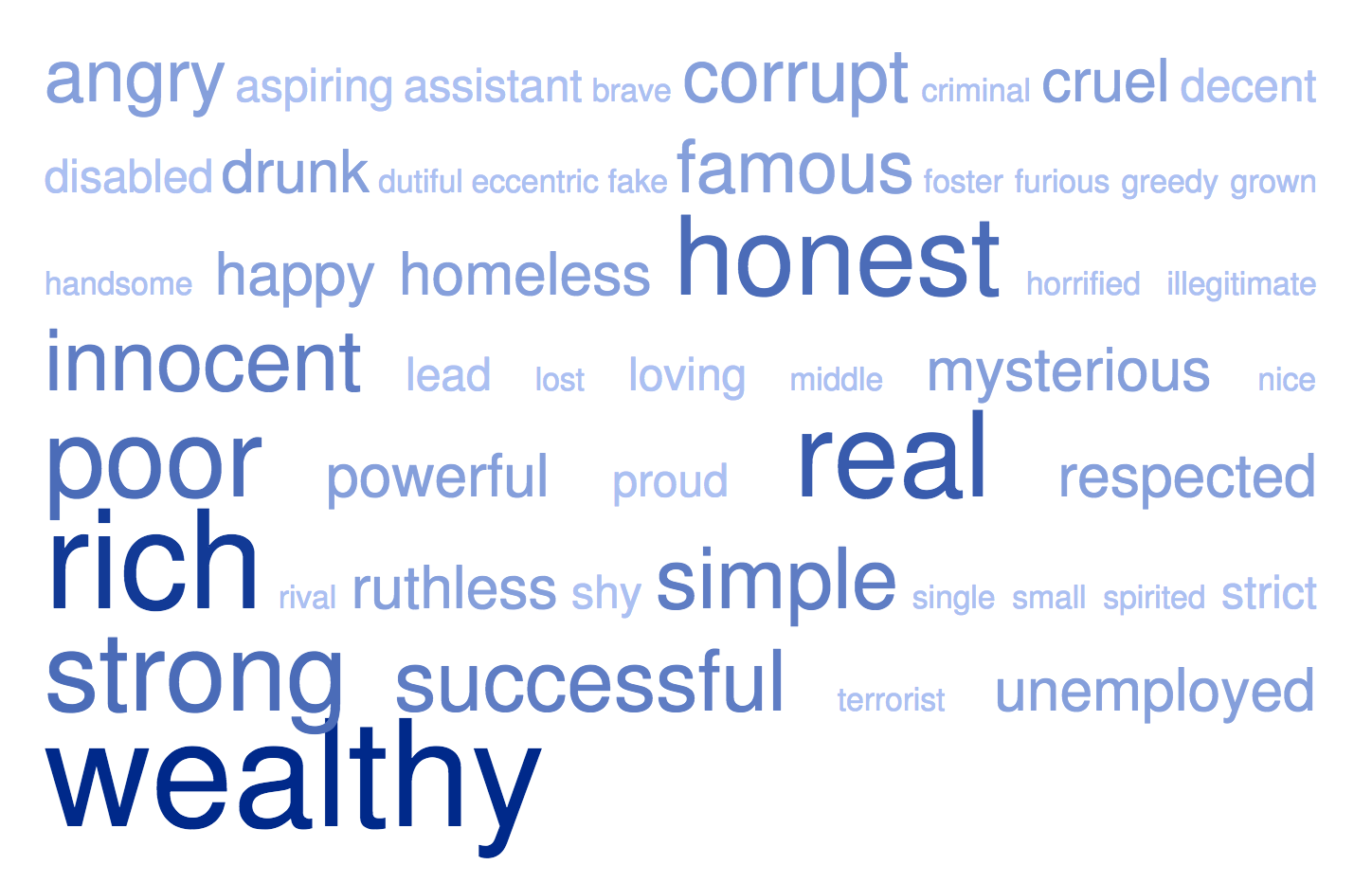}
\end{subfigure}\hspace*{\fill}
\begin{subfigure}{0.25\textwidth}
 \includegraphics[width=\linewidth]{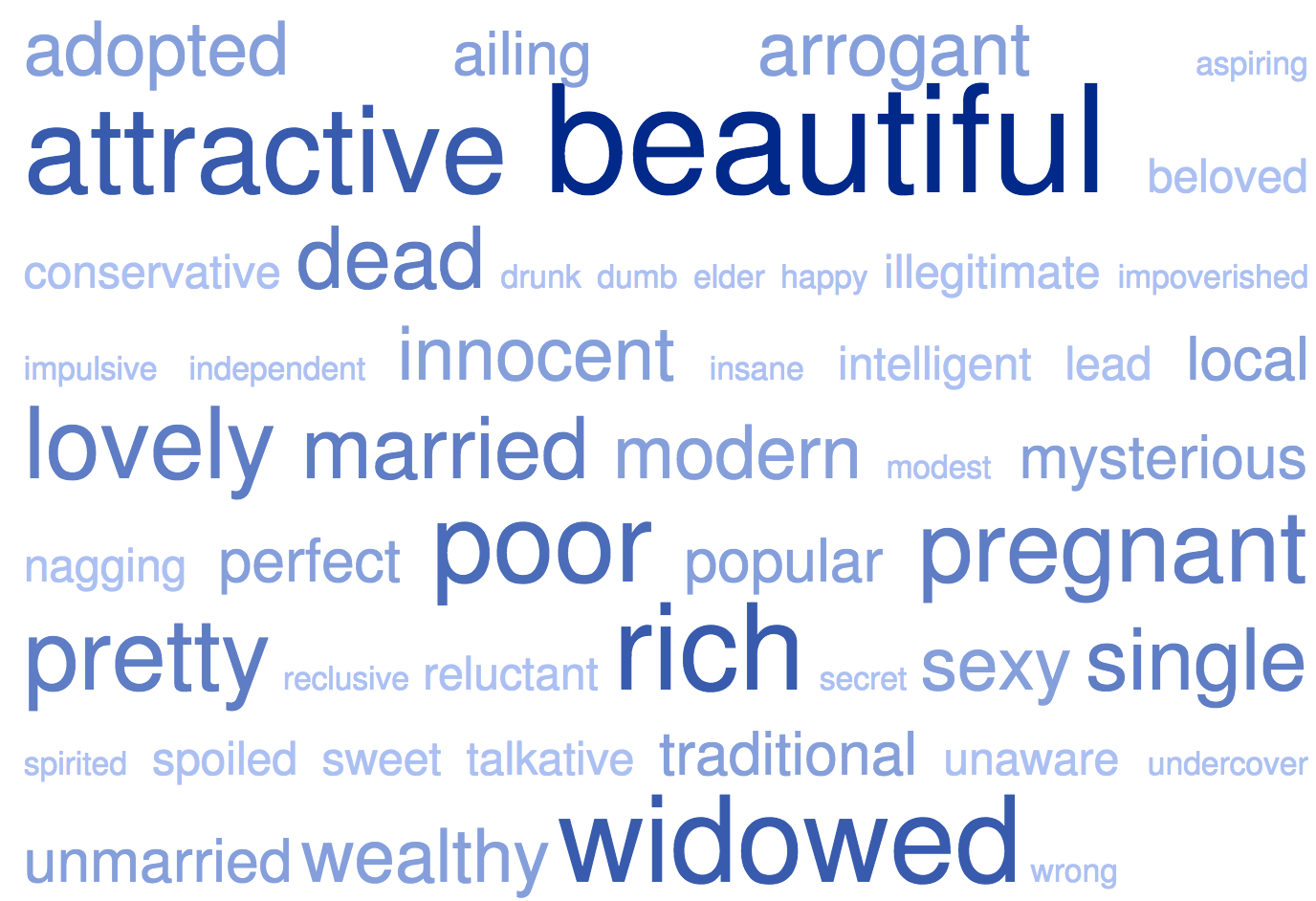}
\end{subfigure}\hspace*{\fill}
    \caption{Adjectives used with males and females}
    \label{fig:adj}
\end{figure}

% Column-wise Verbs
\begin{figure}[t!]
\begin{subfigure}{0.25\textwidth}
 \includegraphics[width=\linewidth]{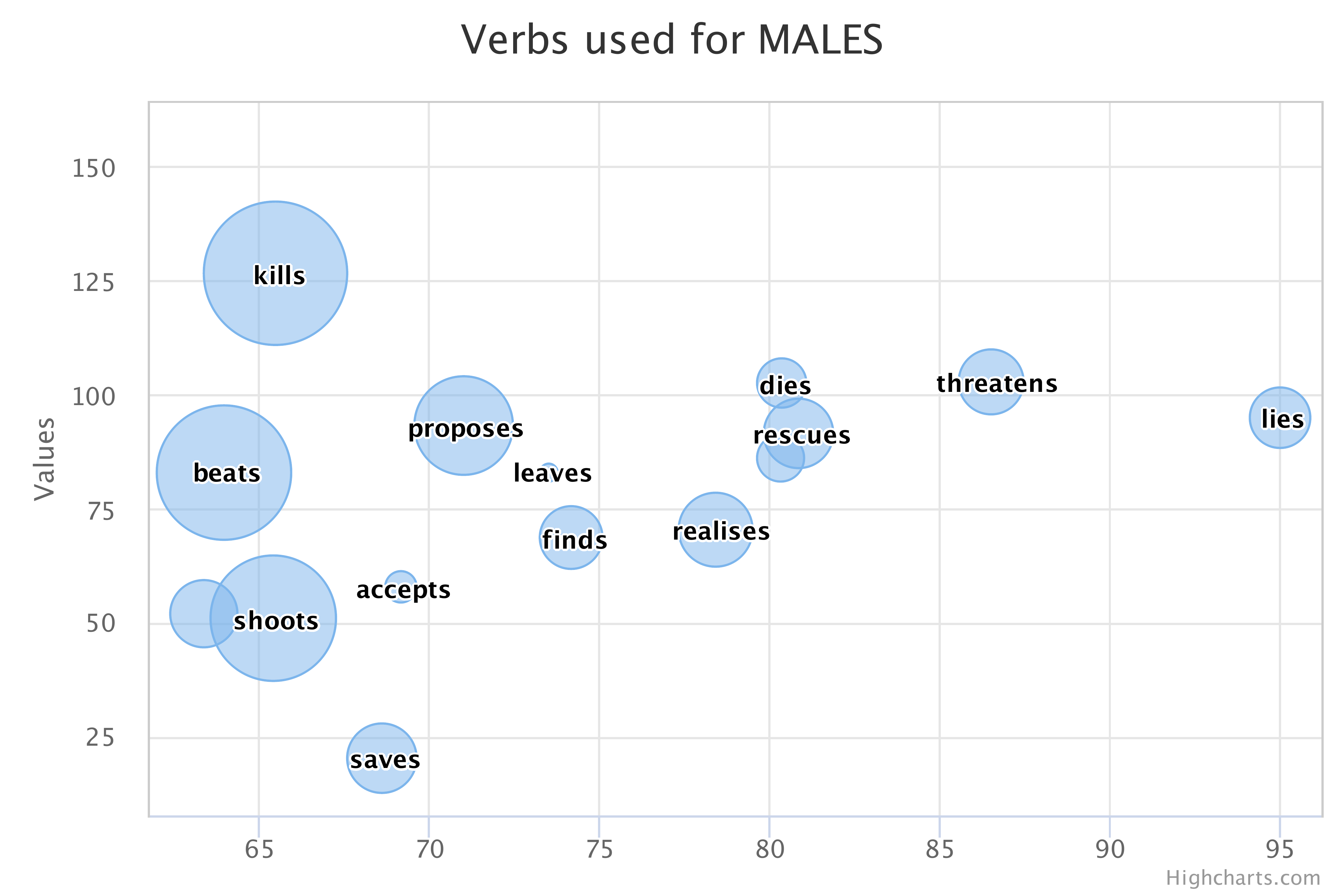}
\end{subfigure}\hspace*{\fill}
\begin{subfigure}{0.25\textwidth}
 \includegraphics[width=\linewidth]{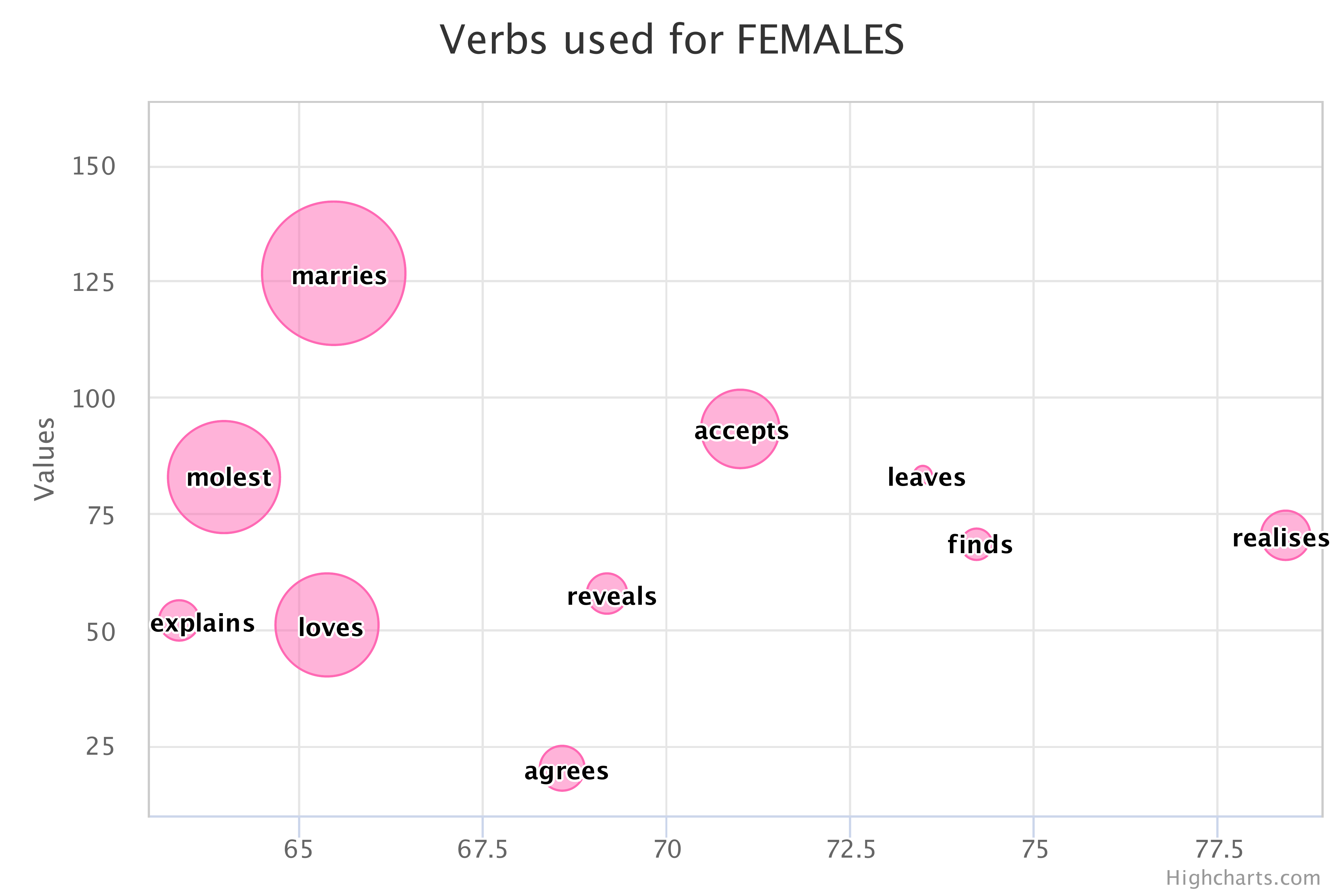}
\end{subfigure}\hspace*{\fill}
    \caption{Verbs used with males and females}
    \label{fig:verbs}
\end{figure}
\begin{figure}[t!]
\begin{subfigure}{0.25\textwidth}
 \includegraphics[width=\linewidth]{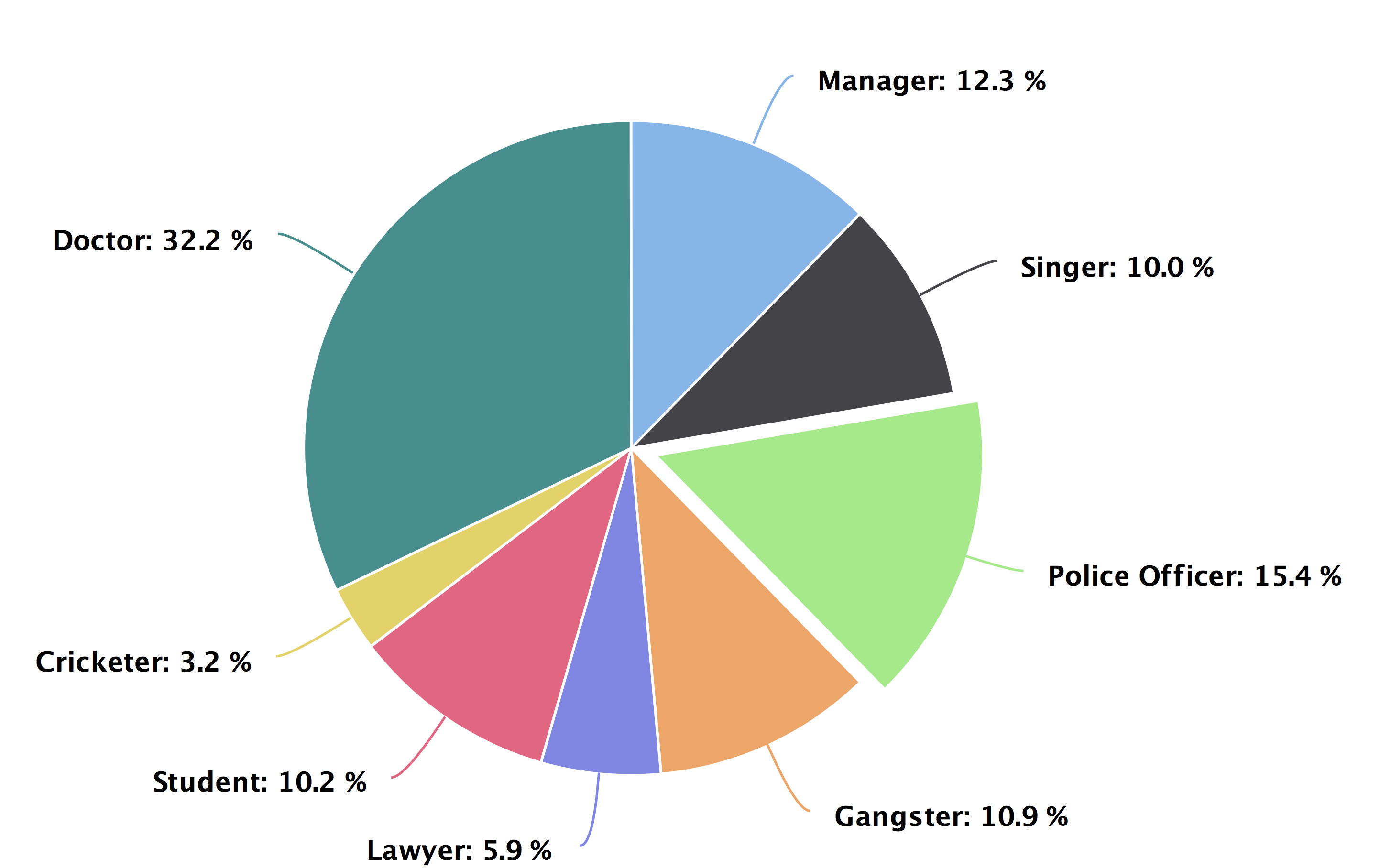}
\end{subfigure}\hspace*{\fill}
\begin{subfigure}{0.25\textwidth}
 \includegraphics[width=\linewidth]{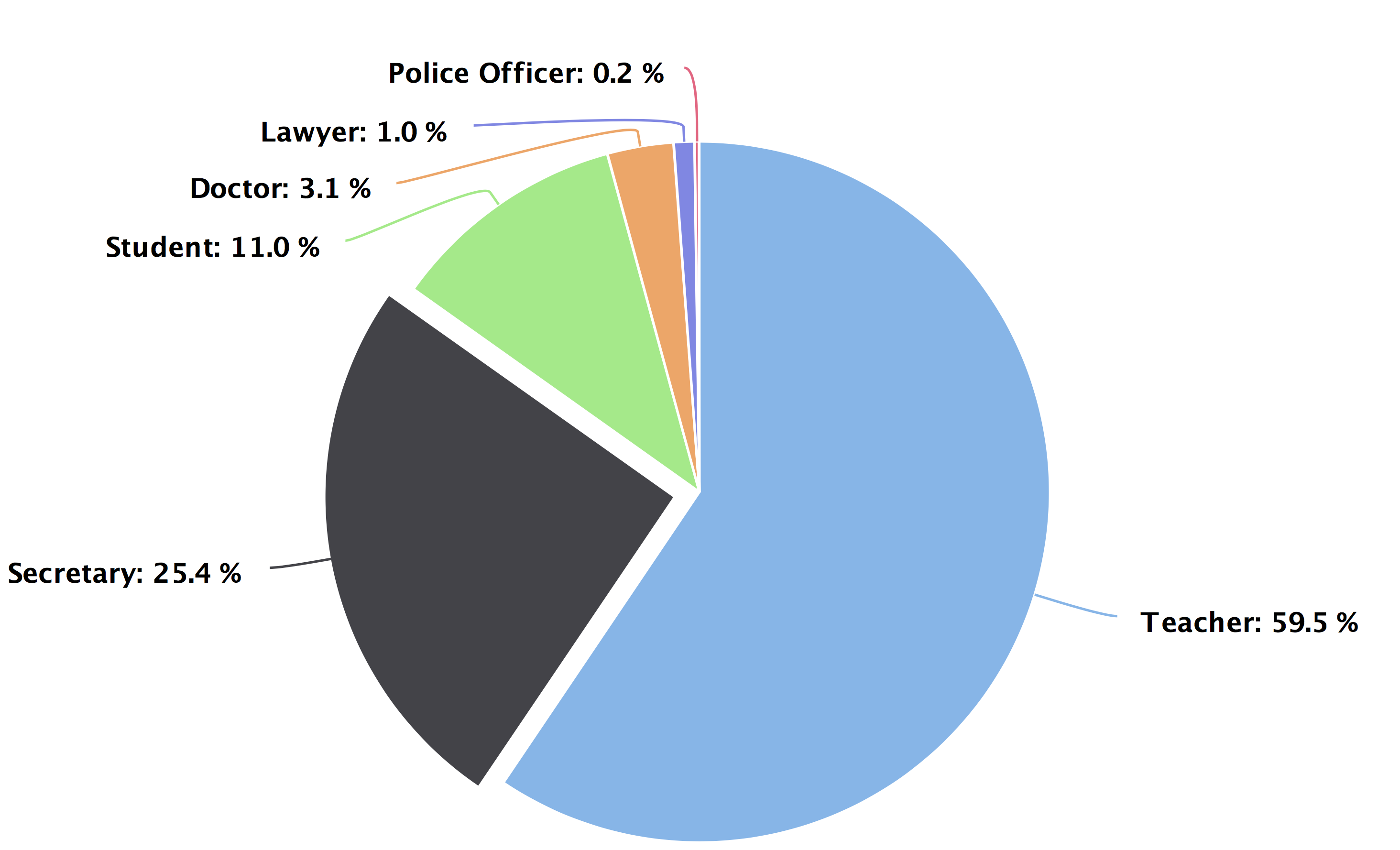}
\end{subfigure}\hspace*{\fill}
    \caption{Occupations of males and females}
    \label{fig:occupations}
\end{figure}

\begin{figure}[t!]
 \includegraphics[width=1.0\columnwidth]{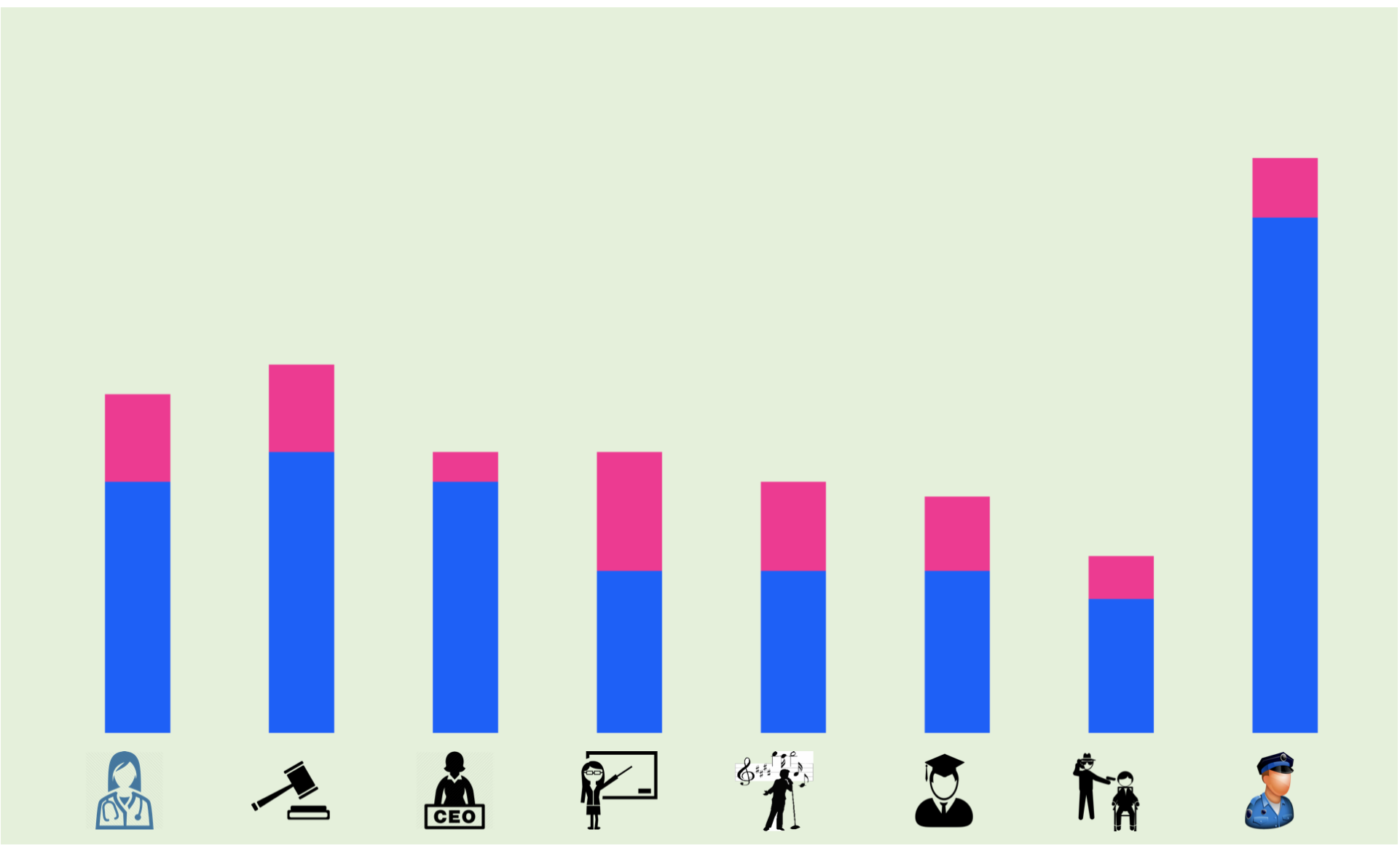}
  \caption{Gender-wise Occupations in Bollywood movies}
  \label{fig:occ}
\end{figure}

To make plots analysis ready, we used OpenIE \cite{fader2011identifying} for performing co-reference resolution on movie plot text. The co-referenced plot is used for all analyses. 
 
The following intra-sentence analysis is performed
\begin{asparaenum}[1)]
    \item \textbf{Cast Mentions in Movie Plot} - We extract mentions of male and female cast in the co-referred plot. The motivation to find mentions is how many times males have been referred to in the plot versus how many times females have been referred to in the plot. This helps us identify if the actress has an important role in the movie or not. In Figure \ref{fig:mentions} it is observed that, a male is mentioned around 30 times in a plot while a female is mentioned only around 15 times. {\it Moreover, there is a consistency of this ratio from 1970 to 2017(for almost 50 years)!}
 
    \item \textbf{Cast Appearance in Movie Plot} - We analyze how male cast and female cast have been addressed. This essentially involves extracting verbs and adjectives associated with male cast and female cast. To extract verbs and adjectives linked to a particular cast, we use Stanford Dependency Parser \cite{de2006generating}. In Fig \ref{fig:adj} and \ref{fig:verbs} we present the adjectives and verbs associated with males and females. 
    We observe that, verbs like \textit{kills, shoots} occur with males while verbs like \textit{marries, loves} are associated with females. Also when we look at adjectives, males are often represented as rich and wealthy while females are represented as beautiful and attractive in movie plots. 
    
    \item \textbf{Cast Introductions in Movie Plot} - We analyze how male cast and female cast have been introduced in the plot. We use OpenIE \cite{fader2011identifying} to capture such introductions by extracting relations corresponding to a cast. Finally, on aggregating the relations by gender, we find that males are generally introduced with a profession like as {\it a famous singer, an honest police officer, a successful scientist and so on} while females are either introduced using physical appearance like {\it beautiful, simple looking} or in relation to another (male) character (daughter, sister of). The results show that females are always associated with a successful male and are not portrayed as independent while males are portrayed to be successful. 
 
    \item \textbf{Occupation as a stereotype} - We perform a study on how occupations of males and females are represented. To perform this analysis, we collated an occupation list from multiple sources over the web comprising of ~350 occupations. We then extracted an associated "noun" tag attached with cast member of the movie using Stanford Dependency Parser \cite{de2006generating} which is later matched to the available occupation list. In this way, we extract occupations for each cast member. We group these occupations for male and female cast members for all the collated movies. Figure \ref{fig:occupations} shows the occupation distribution of males and females. From the figure it is clearly evident that, males are given higher level occupations than females. Figure \ref{fig:occ} presents a combined plot of percentages of male and female having the same occupation. This plot shows that when it comes to occupation like "teacher" or "student", females are high in number. But for "lawyer" and "doctor" the story is totally opposite.

    \begin{figure}[t!]
 \includegraphics[width=1.0\columnwidth]{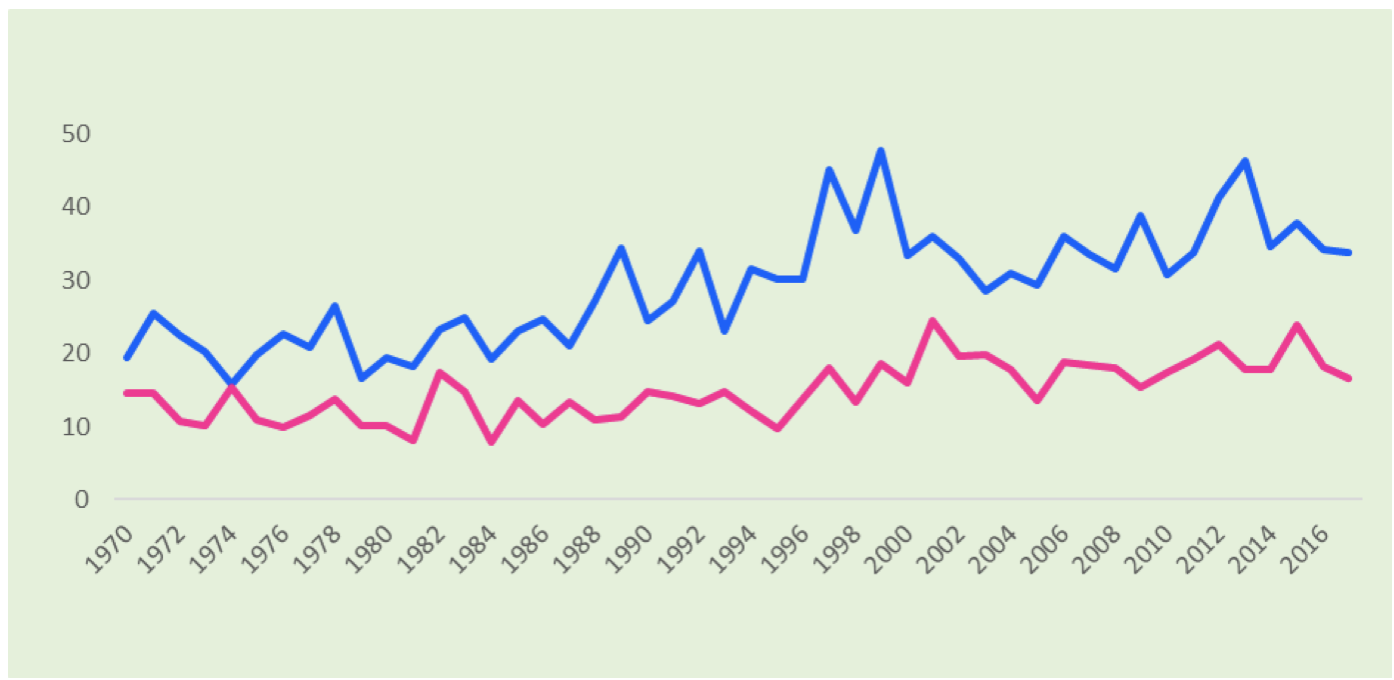}
  \caption{Total Cast Mentions showing mentions of male and female cast. Female mentions are presented in pink and Male mentions in blue}
  \label{fig:mentions}
\end{figure}

     \begin{figure}[t!]
 \includegraphics[width=1.0\columnwidth]{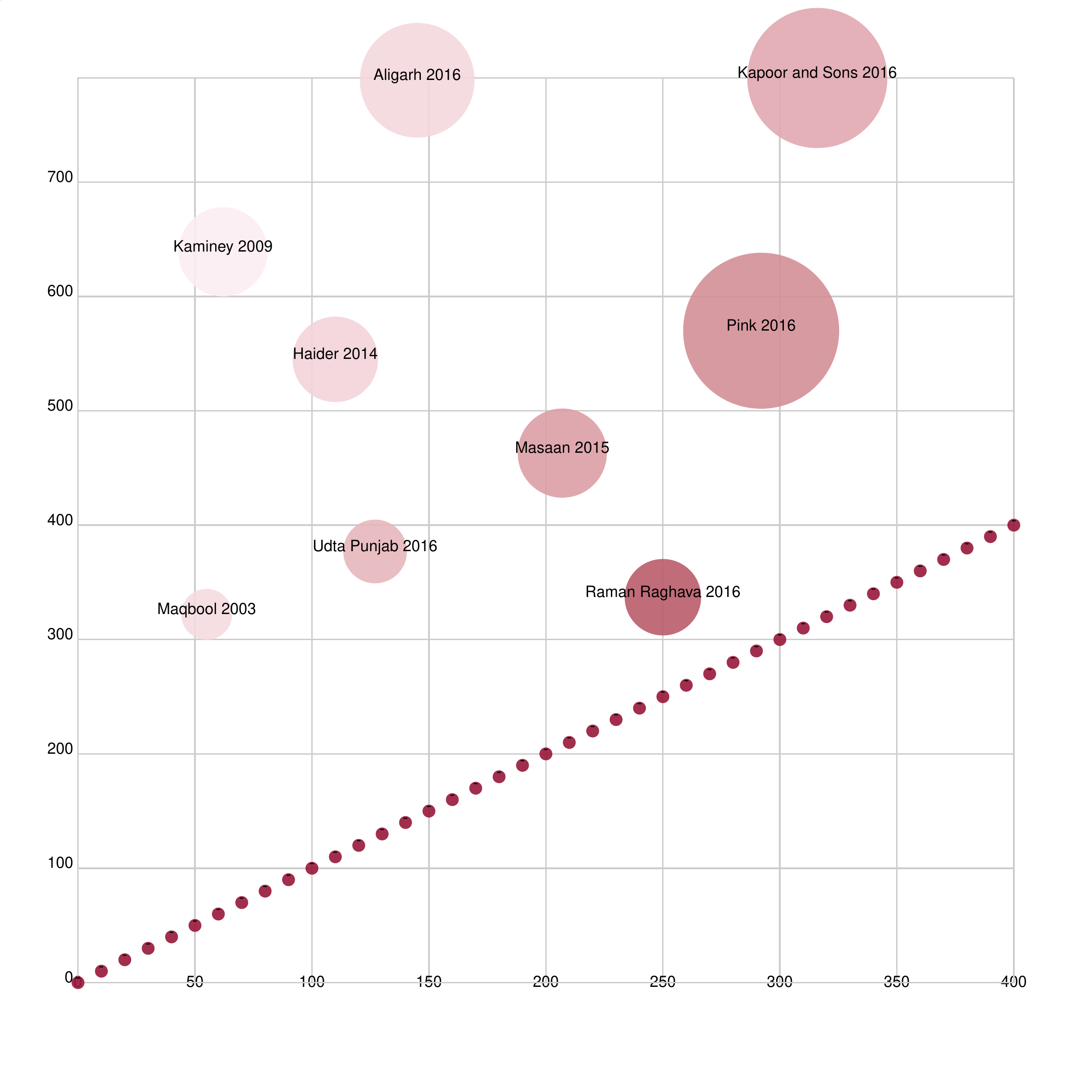}
  \caption{Total Cast dialogues showing ratio of male and female dialogues. Female dialogues are presented on X-axis and Male dialogues on Y-axis }
  \label{fig:castdial}
\end{figure}

\item \textbf{Singers and Gender distribution in Soundtracks} - We perform an analysis on how gender-wise distribution of singers has been varying over the years. To accomplish this, we make use of Soundtracks data present for each movie. This data contains information about songs and their corresponding singers. We extracted genders for each listed singer using their Wikipedia page and then aggregated the numbers of songs sung by males and females over the years. In Figure \ref{fig:soundtrack}, we report the aforementioned distribution for recent years ranging from 2010-2017. We observe that the gender-gap is almost consistent over all these years.

     \begin{figure}[t!]
 \includegraphics[width=1.0\columnwidth]{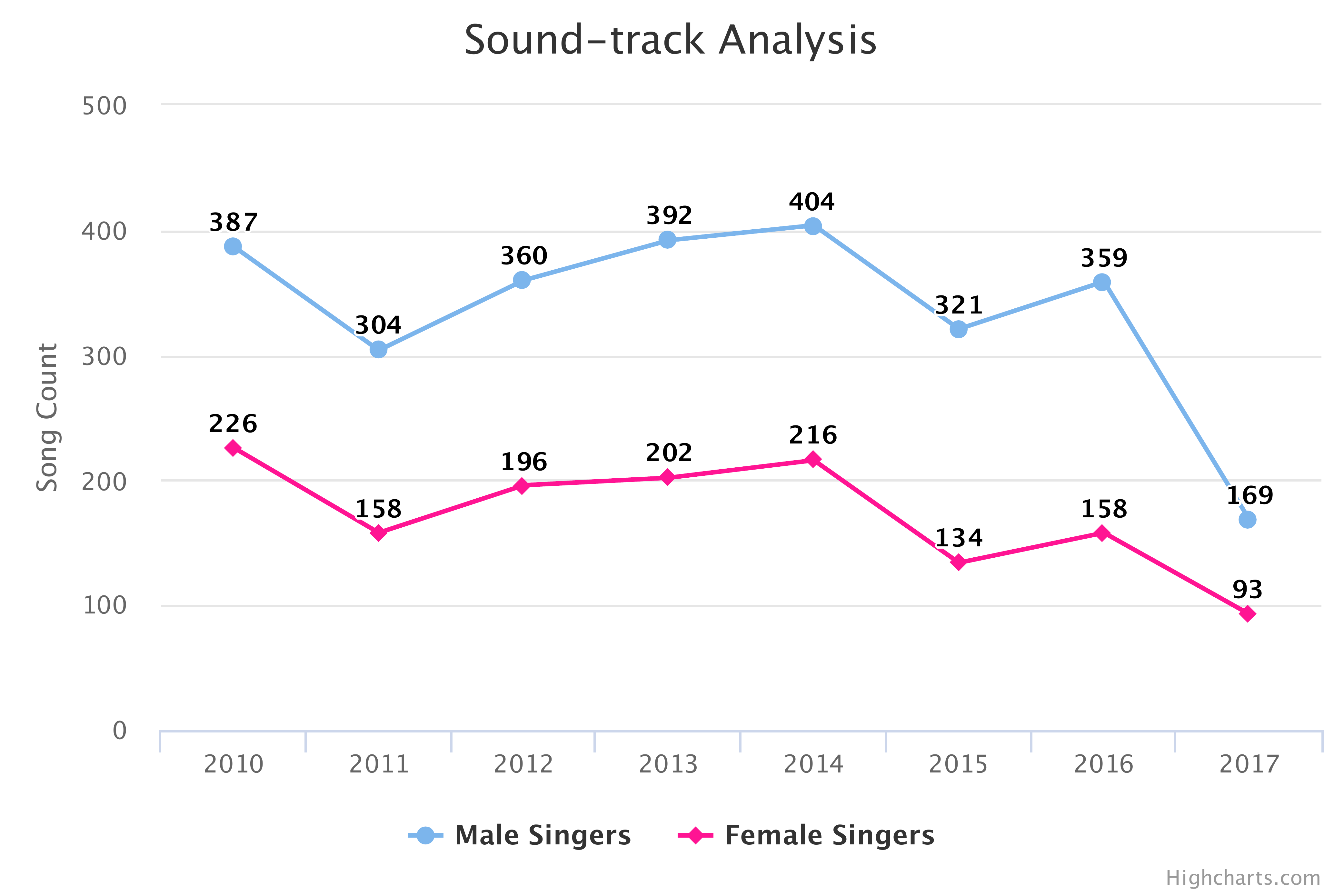}
  \caption{Gender-wise Distribution of singers in Soundtracks}
  \label{fig:soundtrack}
\end{figure}
 Please note that currently this analysis only takes into account the presence or absence of female singer in a song. If one takes into account the actual part of the song sung, this trend will be more dismal. In fact, in a recent interview \footnote {goo.gl/BZWjWG} this particular hypothesis is even supported by some of the top female singers in Bollywood.  In future we plan to use audio based gender detection to further quantify this.

 \item \textbf{Cast Dialogues and Gender Gap in Movie Scripts} - We perform a sentence level analysis on 13 movie scripts available online. We have worked with PDF scripts and extracted structured pieces of information using \cite{ibm} pipeline in the form of structured HTML. We further extract the dialogues for a corresponding cast and later group this information to derive our analysis. 

\begin{figure}[t!]
 \includegraphics[width=1.0\columnwidth]{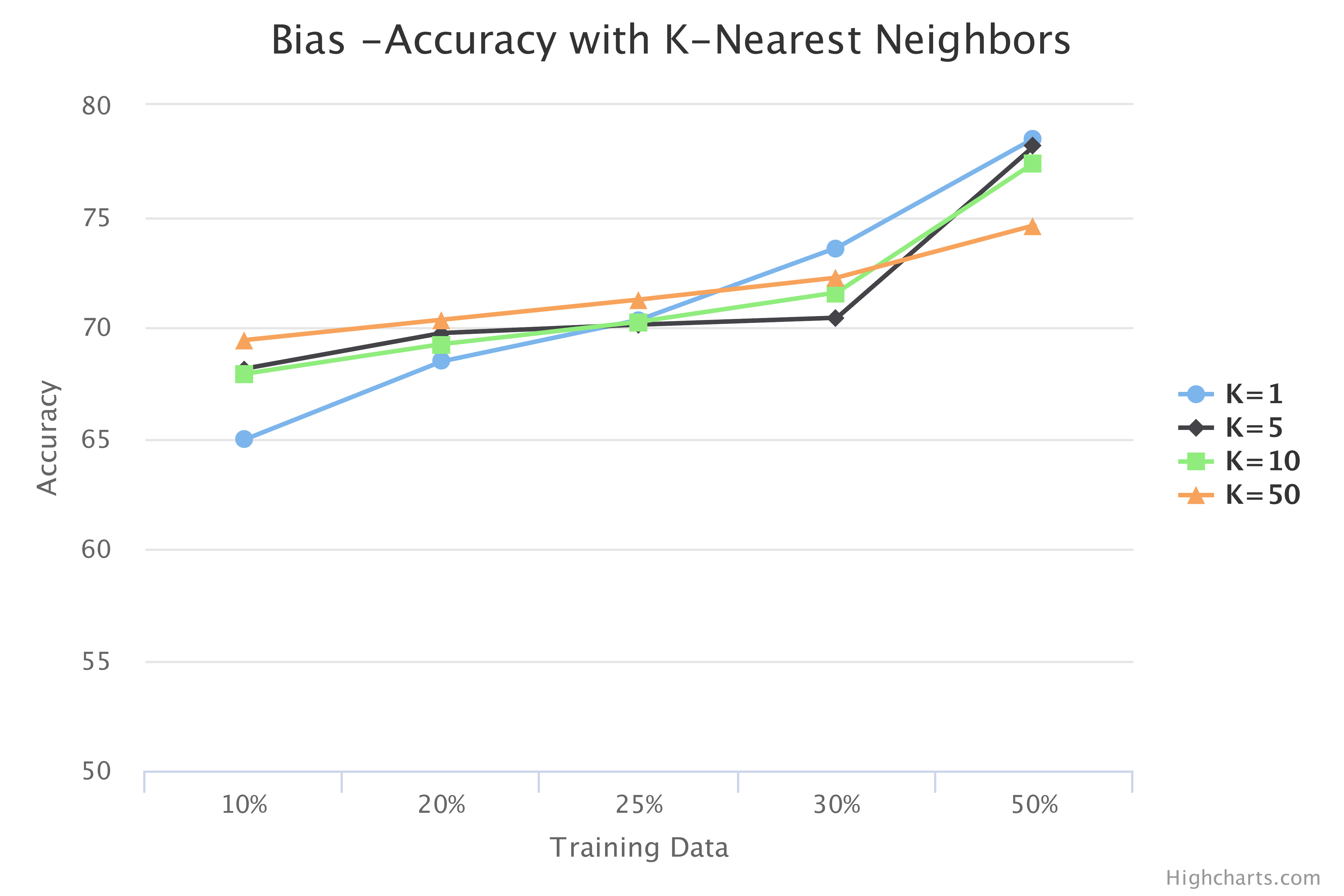}
  \caption{Representing variation of Accuracy with training data}
  \label{fig:bias-knn}
\end{figure}

We first study the ratio of male and female dialogues. In figure \ref{fig:castdial}, we present a distribution of dialogues in males and females among different movies. X-Axis represents number of female dialogues and Y-Axis represents number of male dialogues. The dotted straight line showing $y=x$. Farther a movie is from this line, more biased the movie is. In the figure \ref{fig:castdial}, \textit{Raman Raghav} exhibits least bias as the number of male dialogues and female dialogues distribution is not skewed. As opposed to this, \textit{Kaminey} shows a lot of bias with minimal or no female dialogues.

% \begin{figure}
%  \includegraphics[width=1.0\columnwidth]{illustrations/castdial-lead.png}
%   \caption{Lead Cast dialogues of males and females from different movie scripts}
%   \label{fig:leadcastdial}
% \end{figure}

\begin{figure}
 \includegraphics[width=1.0\columnwidth]{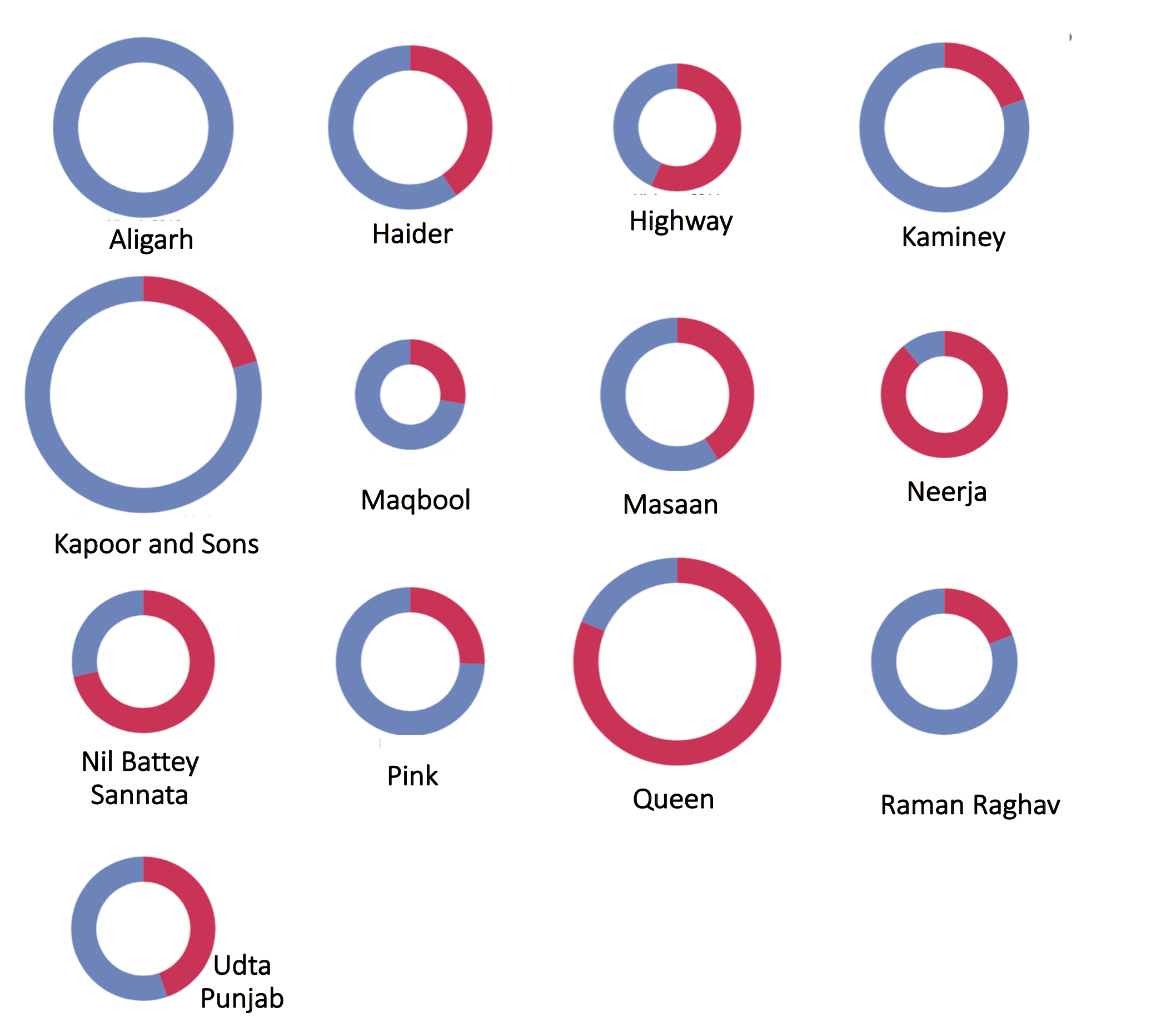}
  \caption{Lead Cast dialogues of males and females from different movie scripts}
  \label{fig:leadcastdial}
\end{figure}

\end{asparaenum}
\subsubsection{Tasks at Inter-Sentence level}
% Column wise KG
\begin{figure}[t!]
\begin{subfigure}{0.25\textwidth}
 \includegraphics[width=\linewidth]{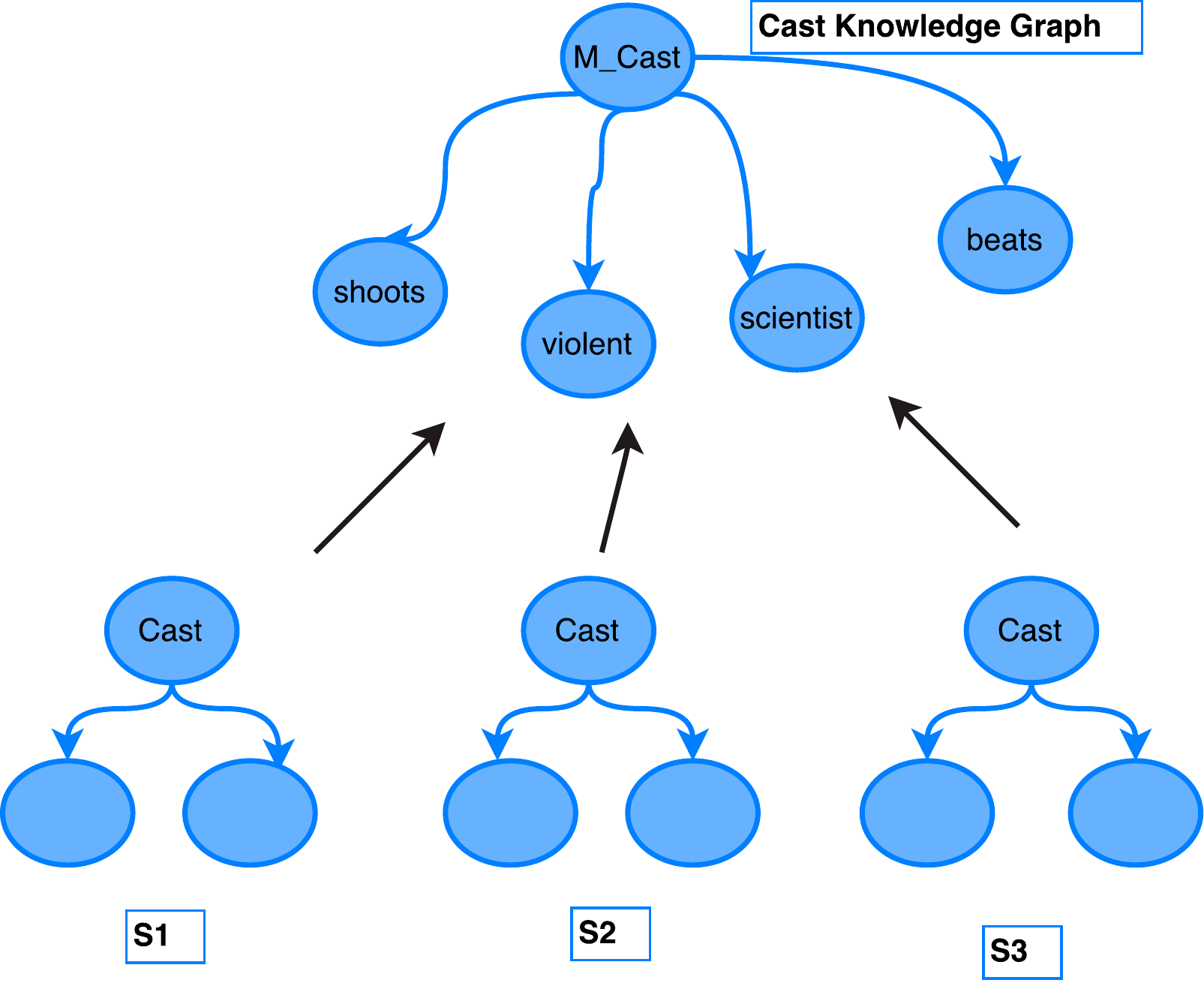}
\end{subfigure}\hspace*{\fill}
\begin{subfigure}{0.25\textwidth}
 \includegraphics[width=\linewidth]{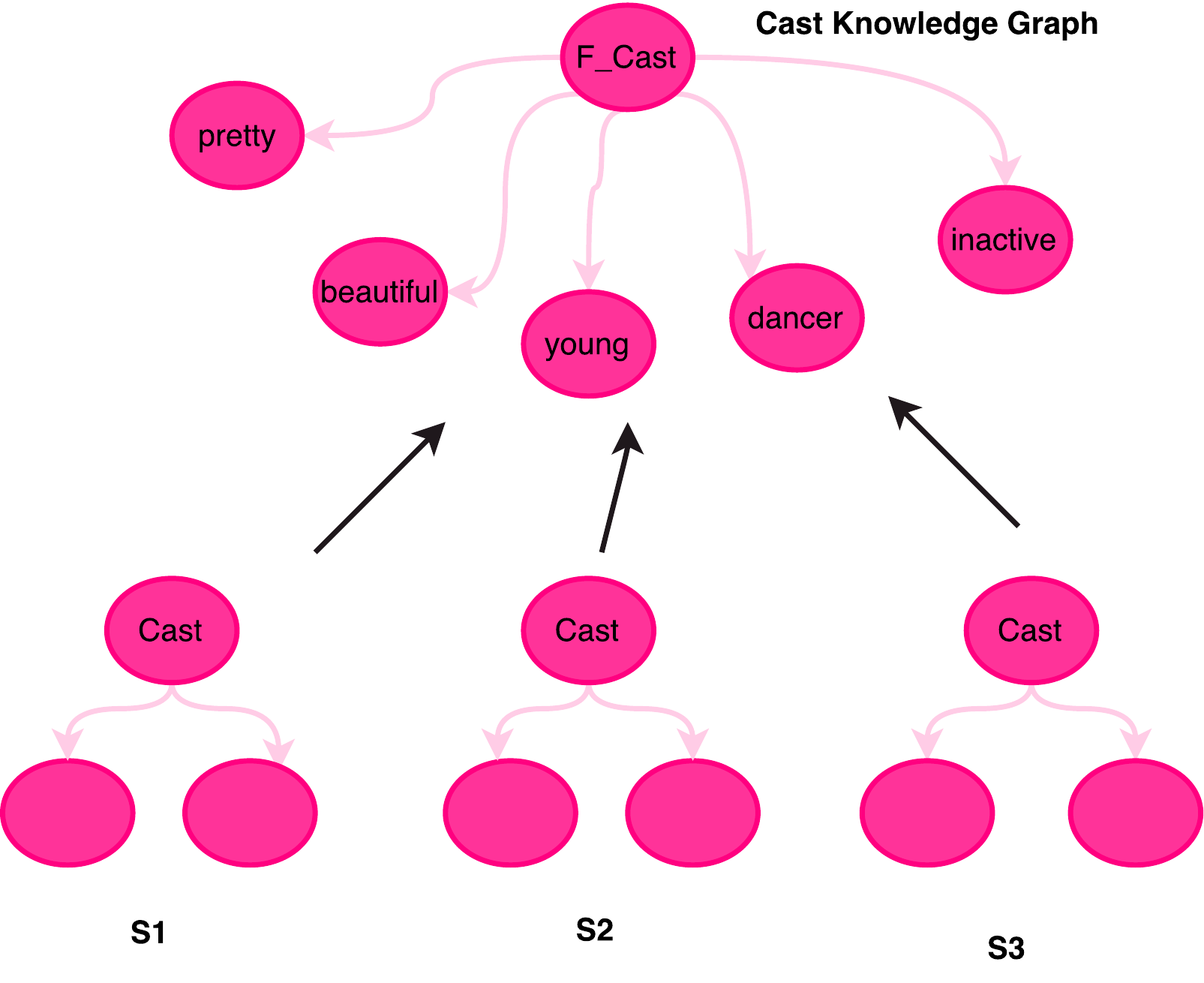}
\end{subfigure}\hspace*{\fill}
    \caption{Knowledge graph for Male and Female Cast}
    \label{fig:kg}
\end{figure}
We analyze the Wikipedia movie data by exploiting plot information. This information is collated at inter-sentence level to generate a context flow using a word graph technique. We construct a word graph for each sentence by treating each word in sentence as a node, and then draw grammatical dependencies extracted using Stanford Dependency Parser \cite{de2006generating} and connect the nodes in the word graph. Then using word graph for a sentence, we derive a knowledge graph for each cast member. The root node of knowledge graph is $[CastGender,CastName]$ and the relations represent the dependencies extracted using dependency parser across all sentences in the movie plot. This derivation is done by performing a merging step where we merge all the existing dependencies of the cast node in all the word graphs of individual sentences. Figure \ref{fig:kg} represents a sample knowledge graph constructed using individual dependencies.

% Then we merge $v\in V_{s_1}$ where $v\in CastSet$ and  $v'\in V_{s_2}$ where $v'\in CastSet$ and $V_{s_1}, V_{s_2}$ represent vertex sets for two sentences $s_1 and s_2$ taken from plot text.

% We perform this merging step for all cast vertices in the word graph to get a knowledge graph corresponding to each $c \in CastSet$ where $c \gets v\in V_{s_1}...V_{s_n}$. In figure \ref{fig:kg} we present a sample collated knowledge graph for a male cast node and for a female cast node. As, we observe from Figure \ref{fig:kg}, the root node of knowledge graph is $[CastGender,CastName]$ and the relations represent the dependencies extracted using dependency parser across all sentences in the movie plot. 

\begin{figure}[t!]
 \includegraphics[width=1.0\columnwidth]{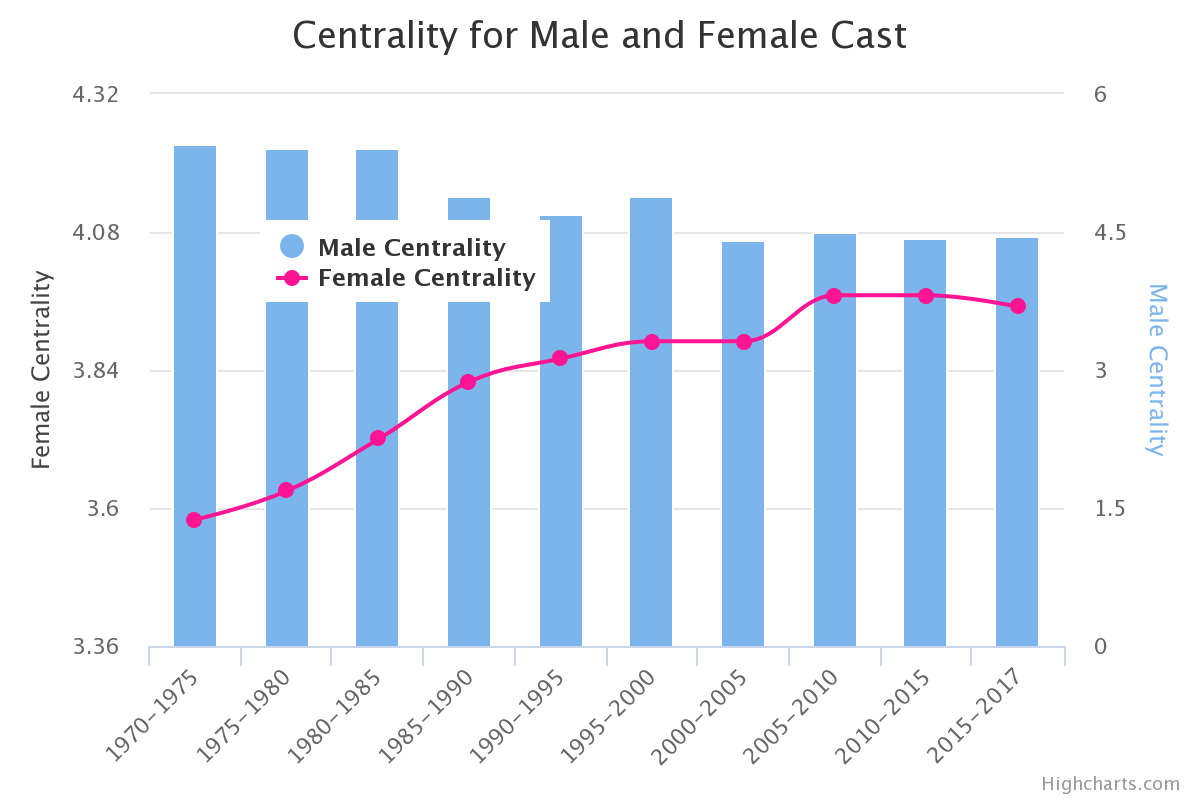}
  \caption{Centrality for Male and Female Cast }
  \label{fig:centrality}
\end{figure}

After obtaining the knowledge graph, we perform the following analysis tasks on the data -

\begin{asparaenum}
    \item \textbf{Centrality of each cast node} - Centrality for a cast is a measure of how much the cast has been focused in the plot. For this task, we calculate between-ness centrality for cast node. Between-ness centrality for a node is number of shortest paths that pass through the node. We find between-ness centrality for male and female cast nodes and analyze the results. In Figure \ref{fig:centrality}, we show male and female centrality trend across different movies over the years. We observe that there is a huge gap in centrality of male and female cast. 

    \item \textbf{Study of bias using word embeddings} - So far, we have looked at verbs, adjectives and relations separately. In this analysis, we want to perform joint modeling of aforementioned. For this analysis, we generated word vectors using Google word2vec \cite{mikolov2013efficient} of length 200 trained on Bollywood Movie data scraped from Wikipedia. CBOW model is used for training Word2vec. The knowledge graph constructed for male and female cast for each movie contains a set of nodes connected to them. These nodes are extracted using dependency parser. We assign a context vector to each cast member node. The context vector consists of average of word vector of its connected nodes. As an instance, if we consider figure \ref{fig:kg}, the context vector for $[M,Cast name]$ would be average of word vectors of (shoots, violent, scientist, beats). In this fashion we assign a context vector to each cast node. 
    
    The main idea behind assigning a context vector is to analyze the differences between contexts for male and female.  
    
    We randomly divide our data into training and testing data. We fit the training data using a K-Nearest Neighbor with varying K. We study the accuracy results by varying samples of train and test data. In Figure \ref{fig:bias-knn}, we show the accuracy values for varying values of K. 
    While studying bias using word embeddings by constructing a context vector, the key point is when training data is 10\%, we get almost 65\%-70\% accuracy, refer to Figure \ref{fig:bias-knn}. This pattern shows very high bias in our data. As we increase the training data, the accuracy also shoots up. There is a distinct demarcation in verbs, adjectives, relations associated with males and females. Although we did an individual analysis for each of the aforementioned intra-sentence level tasks, but the combined inter-sentence level analysis makes the argument of existence of bias stronger. Note the key point is not that the accuracy goes up as the training data is increased. The key point is that since the gender bias is high, the small training data has enough information to classify correctly 60-70\% of the cases.
    
\end{asparaenum}

\subsubsection{Movie Poster and Plot Mentions}
We analyze images on Wikipedia movie pages for presence of males and females on publicity posters for the movie. We use Dense CAP \cite{johnson2016densecap} to extract male and female occurrences by checking our results in the top 5 responses having a positive confidence score. 

After the male and female extraction from posters, we analyze the male and female mentions from the movie plot and co-relate them. The intent of this analysis is to learn how publicizing a movie is biased towards a female on advertising material like posters, and have a small or inconsequential role in the movie.

% \begin{figure*}[t!]
% \begin{subfigure}{1.0\textwidth}
%  \includegraphics[width=\linewidth]{illustrations/Poster.png}
% \end{subfigure}\hspace*{\fill}
%     \caption{Image vs Plot Mentions}
%     \label{fig:poster}
% \end{figure*}

\begin{figure}[t!]
 \includegraphics[width=1.0\columnwidth]{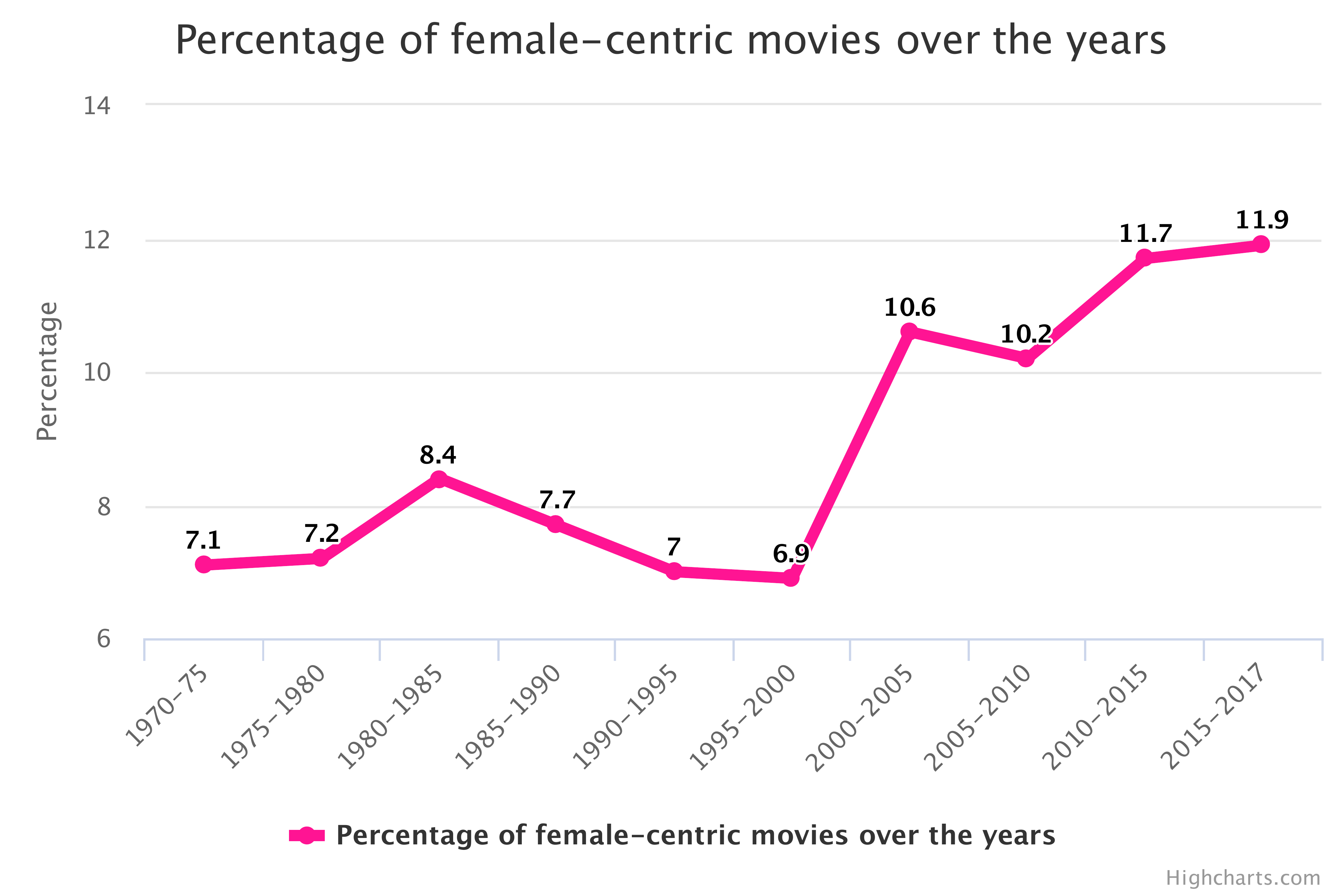}
  \caption{Percentage of female-centric movies over the years }
  \label{fig:percentfc}
\end{figure}

\begin{figure}[t!]
 \includegraphics[width=1.0\columnwidth]{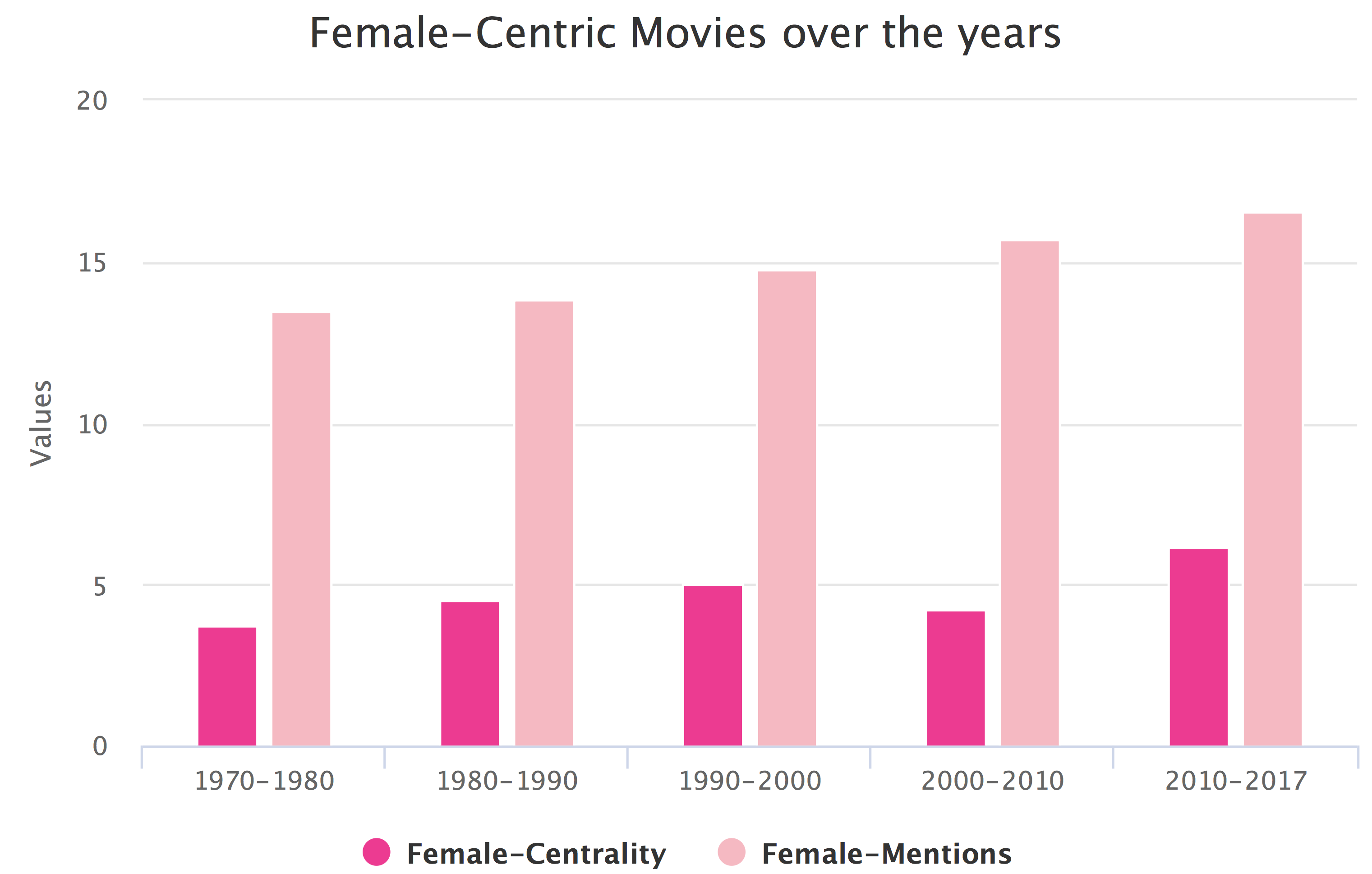}
  \caption{Percentage of female-centric movies over the years }
  \label{fig:fcentre}
\end{figure}

% \begin{figure*}[t!]
% \begin{subfigure}{1.0\textwidth}
%  \includegraphics[width=\linewidth]{illustrations/female-centric-v2.pdf}
% \end{subfigure}\hspace*{\fill}
%     \caption{Female-Centric Movies}
%     \label{fig:fcentre}
% \end{figure*}

While 80\% of the movie plots have more male mentions than females, surprisingly more than 50\% movie posters feature actresses. Movies like GangaaJal \footnote{https://en.wikipedia.org/wiki/Gangaajal}, Platform \footnote{https://en.wikipedia.org/wiki/Platform\_(1993\_film)}, Raees \footnote{https://en.wikipedia.org/wiki/Raees\_(film)} have almost 100+ male mentions in plot but 0 female mentions whereas in all 3 posters females are shown on posters very prominently. 
% {\bf NEED to choose correct words This, itself, points to portrayal of women for purpose of tantalizing the potential audience.}
Also, when we look at Image and Plot mentions, we observe that in 56\% of the movies, female plot mentions are less than half the male plot mentions while in posters this number is around 30\%. Our system detected 289 female-centric movies, where this stereotype is being broken. To further study this, we plotted centrality of females and their mentions in plots over the years for these 289 movies. Figure \ref{fig:fcentre} shows that both plot mentions and female centrality in the plot exhibit an increasing trend which essentially means that there has been a considerable increase in female roles over the years. 
We also study the number of female-centric movies to the total movies over the years. Figure \ref{fig:percentfc} shows the percentage chart and the trend for percentage of female-centric movies. It is enlightening to see that the percentage shows a rising trend. 
Our system discovered at least 30 movies in last three years where females play central role in plot as well as in posters. We also note that over time such biases are decreasing - still far away from being neutral but the trend is encouraging. Figure \ref{fig:percentfc} shows percentage of movies in each decade where women play more central role than male.

% \begin{figure*}[t!]
% \begin{subfigure}{1.0\textwidth}
%  \includegraphics[width=\linewidth]{illustrations/Female-Centric.png}
% \end{subfigure}\hspace*{\fill}
%     \caption{Female-Centric Movies}
%     \label{fig:fcentre}
% \end{figure*}

\subsection{Movie Preview Analysis}
We analyze all the frames extracted from the movie preview dataset and obtain information regarding the presence/absence of a male/female in the frame. If any person is present in the frame we then find out the emotion displayed by the person. The emotion displayed can be one of {angry, disgust, fear, happy, neutral, sad, surprise}. Note that there can be more than one person detected in a single frame, in that instance, emotions of each person is detected. We then aggregate the results to analyze the following tasks on the data - 

\begin{asparaenum}

\item Screen-On Time - Figure \ref{fig:screenontime} shows the percentage distribution of screen-on time for males and female characters in movie trailers. We see a consistent trend across the 10 years where mean screen-on time for females is only a meagre 31.5 \% compared to 68.5 \% of the time for male characters.

\begin{figure}[t!]
\includegraphics[width=1.0\columnwidth]{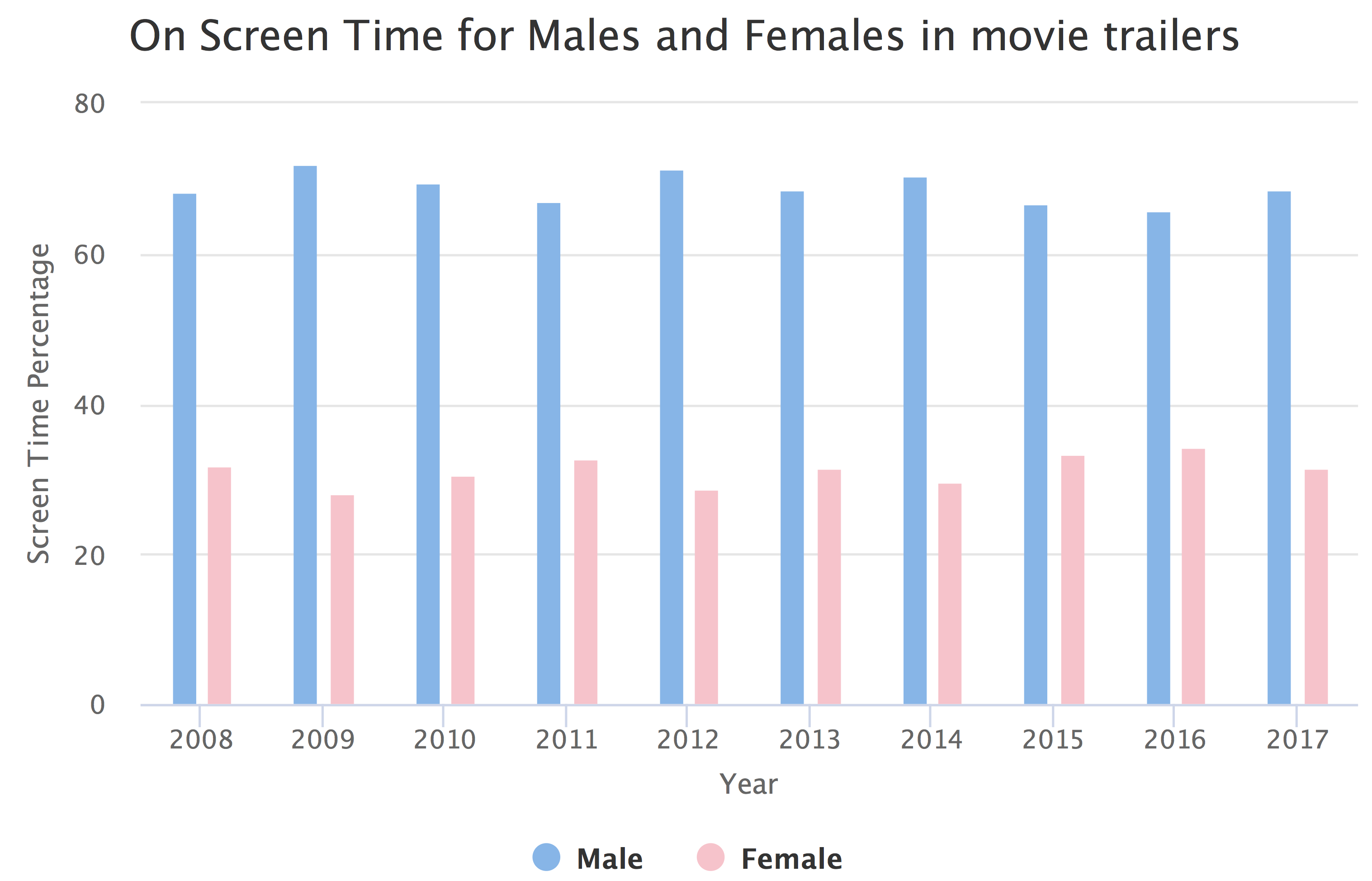}
  \caption{Percentage of screen-on time for males and females over the years}
  \label{fig:screenontime}
\end{figure}

\item Portrayal through Emotions - In this task we analyze the emotions most commonly exhibited by male and female characters in movie trailers. The most substantial difference is seen with respect to the "Anger" emotion. Over the 10 years, anger constitutes 26.3 \% of the emotions displayed by male characters as compared to the 14.5 \% of emotions displayed by female characters. Another trend which is observed, is that, female characters have always been shown as more happy than male characters every year. 
These results correspond to the gender stereotypes which exist in our society. We have not shown plots for other emotions because we could not see any proper trend exhibited by them.

\begin{figure}[t!]
\includegraphics[width=1.0\columnwidth]{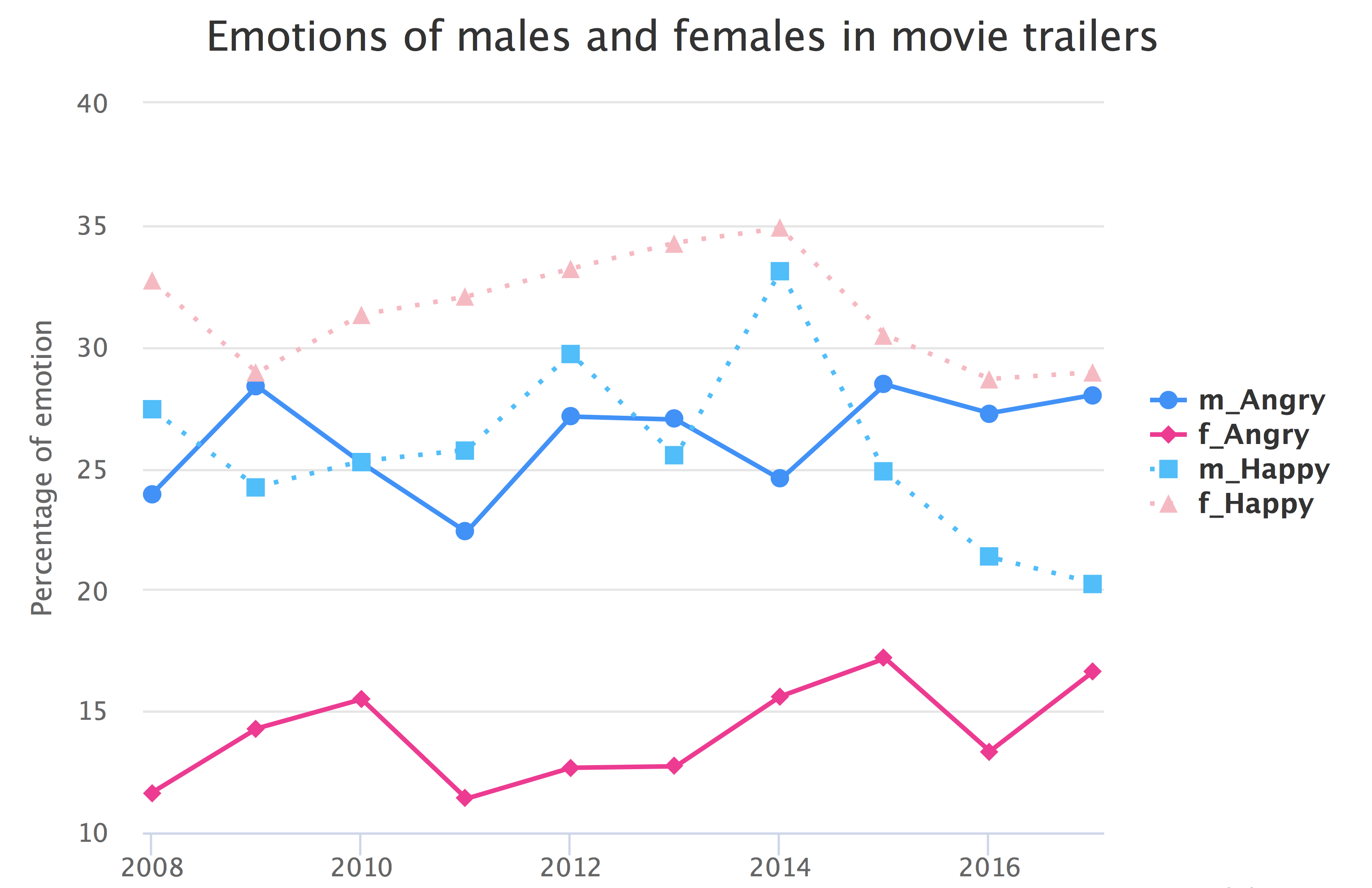}
  \caption{Year wise distribution of emotions displayed by males and females. }
  \label{fig:yearwisedistrutionofemotionsmalesfemales}
\end{figure}

\end{asparaenum}

% \begin{figure*}[t!]
% \begin{subfigure}{1.0\textwidth}
%  \includegraphics[width=\linewidth]{illustrations/Female-Centric.png}
% \end{subfigure}\hspace*{\fill}
%     \caption{Female-Centric Movies}
%     \label{fig:fcentre}
% \end{figure*}

\section{Dataset Release}

We make the dataset accumulated for this analysis public.
\begin {asparaenum}[1)]
\item Movies Dataset - We extracted data for ~4000 movies with cast, soundtracks and plot text. Using OpenIE we generated co-referenced plots. This is located here
%https://github.com/BollywoodMovieDataAAAI/Bollywood-Movie-Data
\footnote{https://github.com/BollywoodMovieDataAAAI/Bollywood-Movie-Data/blob/master/wikipedia-data/coref\_plot.csv}. 
Also, we extracted verbs, relations and adjectives corresponding to each cast and their gender. This information can be found here \footnote{https://github.com/BollywoodMovieDataAAAI/Bollywood-Movie-Data/blob/master/wikipedia-data/}.

\item Movie Preview Dataset - We extracted 880 movie trailers from YouTube from 2008 and 2017. Further, we extracted frames from each trailer which can be found here \footnote{https://github.com/BollywoodMovieDataAAAI/Bollywood-Movie-Data/tree/master/trailer-data/individual-trailers-data}. 
Using this data we calculated the on-screen time and emotions. The detailed results can be found here \footnote{https://github.com/BollywoodMovieDataAAAI/Bollywood-Movie-Data/tree/master/trailer-data}.

\end{asparaenum}

\section{Discussion and Ongoing Work}

While our analysis points towards the presence of gender bias in Hindi movies, it is gratifying to see that the same analysis was able to discover the slow but steady change in gender stereotypes.

We would also like to point out that the goal of this study is not to criticize one particular domain. Gender bias is pervasive in all walks of life including but not limited to the Entertainment Industry, Technology Companies, Manufacturing Factories \& Academia. In many cases, the bias is so deep rooted that it has become the norm. We truly believe that the majority of people displaying gender bias do it unconsciously. We hope that ours and more such studies will help people realize when such biases start to influence every day activities, communications \& writings in an unconscious manner, and take corrective actions to rectify the same.
Towards that goal, we are building a system which can re-write stories in a gender neutral fashion. To start with we are focusing on two tasks:
\begin {asparaenum}[a)]
\item \textbf{Removing Occupation Hierarchy} : It is common in movies, novel \& pictorial depiction to show man as boss, doctor, pilot and women as  secretary, nurse and stewardess. In this work, we presented occupation detection. We are extending this to understand hierarchy and then evaluate if changing genders makes sense or not. For example, while interchanging (\{male, doctor\}, \{female, nurse\}) to (\{male, nurse\}, \{female, doctor\}) makes sense but interchanging \{male, gangster\} to \{female, gangster\} may be a bit unrealistic.
\item \textbf{Removing Gender Bias from plots}: The question we are trying to answer is "If one interchanges all males and females, is the plot/story still possible or plausible?". For example, consider a line in plot  "She gave birth to twins", of course changing this from she to he leads to impossibility. Similarly, there could be possible scenarios but may be implausible like the gangster example in previous paragraph.
\end{asparaenum}
Solving these problems would require development of novel text algorithms, ontology construction, fact (possibility) checkers and implausibility checkers. We believe it presents a challenging research agenda while drawing attention to an important societal problem.

\section{Conclusion}
This paper presents an analysis study which aims to extract existing gender stereotypes and biases from Wikipedia Bollywood movie data containing ~4000 movies. The analysis is performed at sentence at multi-sentence level and uses word embeddings by adding context vector and studying the bias in data. We observed that while analyzing occupations for males and females, higher level roles are designated to males while lower level roles are designated to females. A similar trend has been exhibited for \textit{centrality} where females were less central in the plot vs their male counterparts. Also, while predicting gender using context word vectors, with very small training data, a very high accuracy is observed in gender prediction for test data reflecting a substantial amount of bias present in the data. We use this rich information extracted from Wikipedia movies to study the dynamics of the data and to further define new ways of removing such biases present in the data.
As a part of future work, we aim to extract summaries from this data which are bias-free. In this way, the next generations would stop inheriting bias from previous generations. While the existence of gender bias and stereotype is experienced by viewers of Hindi movies, to the best of our knowledge this is first study to use computational tools to quantify and trend such biases. 

% \section{Acknowledgement}
% The authors would like to point that all analysis and views presented in the paper are personal opinions of the authors and may not reflect the position of respective employers.
\bibliography{aaai.bib}
\bibliographystyle{aaai}
\end{document}